\documentclass[UTF8, twocolumn, times]{aastex63}
\hypersetup{linkcolor=magenta,citecolor=cyan,filecolor=green,urlcolor=blue}

\usepackage{graphicx}	
\usepackage{amsmath}	
\usepackage{amssymb}	

\usepackage{epsfig}
\usepackage{multirow}
\usepackage{float}
\usepackage{footmisc}

\usepackage{enumitem}
\setlist[itemize]{leftmargin=*}

\def\mnras{MNRAS}
\def\apj{ApJ}

\def\apjl{ApJL}
\def\apjs{ApJS}
\def\aap{A\& A}

\def\pasj{PASJ}
\def\nat{Nature}

\def\xmm{{\it XMM-Newton}}
\def\cxo{{\it Chandra}}
\def\swift{{\it Swift}}
\def\nustar{{\it NuSTAR}}

\def\sw16{Swift J1644+57}
\def\asa{ASASSN-14li}
\def\igr{IGR J12580+0134}

\submitjournal{ApJ}

\shorttitle{Short-term X-ray Variability of the TDE \sw16}
\shortauthors{Jin C.}

\begin{document}

\title{Long-term Evolution of the Short-term X-ray Variability of the Jetted TDE \sw16}

\correspondingauthor{Chichuan Jin}
\email{ccjin@bao.ac.cn}

\author[0000-0002-2006-1615]{Chichuan Jin}
\affiliation{National Astronomical Observatories, Chinese Academy of Sciences, 20A Datun Road, Beijing 100101, China}
\affiliation{School of Astronomy and Space Sciences, University of Chinese Academy of Sciences, 19A Yuquan Road, Beijing 100049, China}



\begin{abstract}
The short-term X-ray variability of tidal disruption events (TDEs) and its similarities with active galactic nuclei (AGN) are poorly understood. In this work, we show the diversity of TDE's short-term X-ray variability, and take \sw16\ as an example to study the evolution of various properties related to the short-term X-ray variability, such as the X-ray flux distribution, power spectral density (PSD), rms variability, time lag and coherence spectra. We find that the flux distribution of \sw16\ has a lognormal form in the normal state, but deviates from it significantly in the dipping state, thereby implying different physical mechanisms in the two states. We also find that during the first two \xmm\ observations in the dipping state, \sw16\ exhibited different variability patterns, which are characterized by steeper PSDs and larger rms than the normal state. A significant soft X-ray lag is detected in these two observations, which is $\sim$ 50 s between 0.3-1 keV and 2-10 keV with a high coherence. Using the 2-10 keV rms of $0.10-0.50$, the black hole mass of \sw16\ is estimated to be $(0.6-7.9)\times10^{6}M_{\odot}$, but the variation of rms as the TDE evolves introduces a large uncertainty. Finally, we discuss the value of conducting similar studies on other TDEs, especially in the coming era of time-domain astronomy when a lot more TDEs will be discovered in X-rays promptly. This also heralds a significant increase in the demand for deep follow-up observations of X-ray selected TDEs with X-ray telescopes of large effective areas and long orbital periods.
\end{abstract}



\section{Introduction}
\label{sec-intro}
A tidal disruption event (TDE) is an astrophysical phenomenon when a star approaches a super-massive black hole (SMBH) and gets fully/partially disrupted by the tidal force of the SMBH. It was predicted that such an event should lead to the accretion of stellar debris by the SMBH, thereby producing temporal X-ray emission with typical decay timescales of month to years (e.g. \citealt{Hills.1975}; \citealt{Lacy.1982}; \citealt{Rees.1988}). Indeed, the first TDEs were detected in the X-ray band by the {\it ROSAT} satellite (e.g. \citealt{Bade.1996}; \citealt{Komossa.1999}; \citealt{Grupe.1999}). Since then there have been $\sim$100 TDEs (or TDE candidates)\footnote{https://tde.space/} detected across the entire electro-magnetic waveband (see the most recent reviews by e.g. \citealt{van Velzen.2020}; \citealt{Saxton.2021}; \citealt{Gezari.2021}).

TDEs have been found in both quiescent galaxies (e.g. \citealt{Lin.2018}; \citealt{Li.2020}) and active galaxies (e.g. \citealt{Blanchard.2017}; \citealt{Liu.2020}; \citealt{Neustadt.2020}). One of the main differences between a TDE and an active galactic nuclei (AGN) is the timescale of their evolution (e.g. \citealt{Auchettl.2018}). AGN are generally persistent sources, producing broadband radiation for millions of years or longer (\citealt{Peterson.1997}). In comparison, a TDE is a transient phenomenon, whose radiation is detectable only for months to years after the initial outburst. The light curve of a TDE generally comprise a rapidly rising phase, which is followed by a slowly decaying phase with a power law index of roughly -5/3 (e.g. \citealt{Rees.1988}; \citealt{Komossa.2015, Komossa.2017}; \citealt{Holoien.2019, Holoien.2020}). However, more complex TDE light curves spanning months to years have also been found in different wavebands (e.g. \citealt{Mangano.2016}; \citealt{Holoien.2018}; \citealt{Shu.2020}; \citealt{Payne.2020}).

The difficulty of detecting a TDE is mainly caused by their unpredictability of position and time, as well as their short duration. For the few dozens of X-ray TDEs reported so far (e.g. \citealt{Komossa.1999}; \citealt{Komossa.2015, Komossa.2017}; \citealt{Auchettl.2017}; \citealt{Saxton.2020}), their discoveries mainly relied on the search of X-ray archive, which has been accumulated during the past few decades by various X-ray satellites such as {\it ROSAT} (\citealt{Truemper.1982}), \xmm\ (\citealt{Jansen.2001}), \cxo\ (\citealt{Weisskopf.2000}) and \swift\ (\citealt{Gehrels.2004}). As a result, these TDEs were often poorly monitored, and are generally too faint to be observed now (see \citealt{Saxton.2021} for a detailed review).

The SMBH in a TDE has a mass of $\lesssim10^{8}~M_{\sun}$ (e.g. \citealt{Mockler.2019}; \citealt{Saxton.2020}), which is consistent with the lower mass range of SMBHs in AGN, such as many narrow line Seyfert 1 galaxies (NLS1: \citealt{Boller.1996}). It is well known that low-mass AGN can exhibit significant short-term\footnote{Throughout the paper, we define {\it short-term} as timescales $\lesssim$ 100 ks, and define {\it long-term} as timescales from days to years.} X-ray variability (e.g. \citealt{Miniutti.2009}; \citealt{Ponti.2012}), which contains crucial information about the SMBH and the physical processes close to its event horizon (see Section~\ref{sec-xray-prop} for more details). Likewise, TDEs can also exhibit short-term X-ray variability (see Figure~\ref{fig-lccompare}). For example, previous studies have discovered that TDEs can exhibit X-ray QPO signals (\citealt{Reis.2012}; \citealt{Pasham.2019}), as well as X-ray time lags consistent with the disc reflection scenario (\citealt{Kara.2016}). Therefore, it is both valuable and feasible to conduct detailed studies of TDE's short-term X-ray variability and, more importantly, to explore its evolution with the TDE phase.

The main idea of this work is to present the first exploratory study of the long-term evolution of various properties related to the short-term X-ray variability of TDEs. This type of study requires a series of deep follow-up observations on X-ray bright TDEs, although such datasets are rare. In this work, we mainly study the famous TDE \sw16, because this source was bright in X-rays and was observed by a series of deep \xmm\ observations (see Table~\ref{tab-obs}). Another two TDEs are adopted for comparison only. \xmm's large effective area and long orbital period make it an ideal instrument to obtain high-quality X-ray data, especially continuous long light curves for the exploration of short-term variability. We will demonstrate that there is a wealth of information to be extracted in the evolution of short-term variability, which can be used to further constrain the X-ray mechanism. Therefore, future studies should pay more attention to the properties embedded in the short-term variability and its long-term evolution.

This paper is structured as follows. Firstly, we explain how the sources and observations are selected for this work, and describe the data reduction procedures. Then in Section 3 we show the diversity of TDE's short-term X-ray light curves. This is followed by Section 4, where we present the study of the evolution of short-term variability properties of \sw16. In Section 5, we discuss how the new results reported for \sw16\ in this work can help to constrain its X-ray mechanism. We also use the rms variability to estimate the black hole mass. Finally, we discuss the great potential and prospects of this type of study, and provide suggestions for future observations. We summarize our work in the final section. 

\begin{table}
\centering
   \caption{\xmm\ observations used in this work.}
    \label{tab-obs}
\begin{tabular}{lcccc}
\hline
\hline
 & Obs-Date & Obs-ID & Duration & $t_{\rm TDE}$ \\
 & & & (ksec) & (days) \\ 
\hline
\multicolumn{5}{l}{\it \sw16: $z=0.3534$} \\
Obs-1 & 2011-03-31 & 0658400701 & 25.7 & 3.4 \\
Obs-2 & 2011-04-16 & 0678380101 & 22.7 & 19.4 \\
Obs-3 & 2011-04-30 & 0678380201 & 27.3 & 33.3 \\ 
Obs-4 & 2011-05-16 & 0678380301 & 27.7 & 49.3 \\ 
Obs-5 & 2011-05-30 & 0678380401 & 27.3 & 63.2 \\ 
Obs-6 & 2011-07-03 & 0678380501 & 17.0 & 97.3 \\ 
Obs-7 & 2011-07-15 & 0678380601 & 27.7 & 109.1 \\ 
Obs-8 & 2011-07-27 & 0678380701 & 16.5 & 121.1 \\ 
Obs-9 & 2011-08-14 & 0678380801 & 27.2 & 139.0 \\ 
Obs-10 & 2011-08-27 & 0678380901 & 26.7 & 153.0 \\ 
Obs-11 & 2011-09-06 & 0678381001 & 26.2 & 163.0 \\ 
Obs-12 & 2011-09-20 & 0678381101 & 19.8 & 176.9 \\ 
Obs-13 & 2011-10-02 & 0678381201 & 26.2 & 188.9 \\ 
\hline
\multicolumn{5}{l}{\it \asa: $z=0.0206$} \\
Obs-1 & 2014-12-08 & 0722480201 & 93.5 & 15.9 \\
Obs-2 & 2015-12-10 & 0770980101 & 95.0 & 382.9 \\
\hline
\multicolumn{5}{l}{\it \igr: $z=0.0041$} \\
Obs-1 & 2011-01-22 & 0658400601 & 19.3 & 20.7 \\
\hline
\end{tabular}
\\
\smallskip
Notes. {\it Duration} is the observing time before GTI correction in EPIC-pn, except Obs-1 of \sw16\ when only MOS1 data were used. $t_{\rm TDE}$ is the number of days since the discovery of the TDE. The first detection date of \igr\ has some uncertainty, and we assume it to be 2011-01-02.
\end{table}

\section{Sample \& Observations}
\label{sec-obs}
\subsection{The TDE Sample and Observations}
\label{sec-sample}
The study of long-term evolution of the short-term X-ray variability has high demands on the dataset. Firstly, it requires the TDE to be bright enough in X-rays and show significant short-term variability. Secondly, it requires a series of follow-up observations with long exposures (i.e. at least several tens of kilo-seconds of continuous exposure per observation). So far very few TDEs can meet these requirements, either because the source was not bright enough, or not variable enough, or there were too few follow-up observations, or those observations were too short. After going through literatures and data archives, we select \sw16\ as the primary source for our case study, and choose another two TDEs for comparison.

\subsubsection{\sw16} 
\sw16\ (Ra=16$^{\rm h}$44$^{\rm m}$49.97$^{\rm s}$, Dec=+57$\degr$34$'$59.7$''$) is an X-ray bright TDE originally discovered by \swift\ on 2011-03-28 as a $\gamma$-ray burst (\citealt{Bloom.2011}; \citealt{Burrows.2011}; \citealt{Levan.2011}; \citealt{Zauderer.2011}). It is located in a galaxy at redshift $z=0.3534$ (\citealt{Yoon.2015}). The X-ray emission of this TDE was bright and long lasting (i.e. hundreds of days: \citealt{Seifina.2017} and references therein). It was suggested that the X-ray emission should originate from a jet which points towards the observer, and so the relativistic beaming effect boosts the X-ray flux (\citealt{Zauderer.2013}; \citealt{Mangano.2016}).

Being a rare jetted TDE, \sw16\ has been well monitored since its discovery. \xmm\ observed it 13 times from 2011-03-31 to 2011-10-02, and each of these observations lasted for more than 16 ks (see Table~\ref{tab-obs}). In the meantime, \swift\ monitored it from 2011-03-28 to 2020-09-11 for 1000 snapshots. Significant X-ray variability has been discovered in these follow-up observations, especially the dips in the long-term X-ray light curve, which are suggested to be the consequence of jet procession and/or nutation (\citealt{Saxton.2012}; \citealt{Lei.2013}). Using \xmm\ and {\it Suzaku} data,  \citet{Reis.2012} discovered a $\sim$200 s quasi-periodic oscillation (QPO) signal in the X-ray light curves of \sw16. \citet{Kara.2016} reported a relativistic reverberation lag of $\sim$100 s in the Iron line region of 6-8 keV. Therefore, \sw16\ offers a good opportunity to study the long-term evolution of the short-term X-ray variability.

We use all the \xmm\ data of \sw16\ to perform the analysis. The \swift\ X-ray Telescope (XRT) data are also used to reveal the long-term evolution of the X-ray flux state.

\subsubsection{\asa\ and \igr}
In order to show the diversity of short-term X-ray variability, we select another two TDEs to compare with \sw16.

The first source is \asa\ (Ra=12$^{\rm h}$48$^{\rm m}$15.23$^{\rm s}$, Dec=+17$\degr$46$'$26.22$''$), which is another famous TDE firstly detected by the All-Sky Automated Survey for SuperNovae (ASAS-SN: \citealt{Shappee.2014}) on 2014-11-22 (\citealt{Jose.2014}). The source is located in the galaxy PGC 043234 at redshift $z=0.0206$ (\citealt{Jose.2014}; \citealt{Holoien.2016}). Being a nearby X-ray bright TDE, \asa\ was observed by \xmm\ 13 times from 2014-12-06 to 2019-06-06, and each of them lasted for 10-97 ks (see Table~\ref{tab-obs}). \swift\ monitored this TDE with 131 observations between 2015-01-03 and 2019-05-13. The X-ray spectrum of \asa\ appears disc-dominated (\citealt{Kara.2018}; \citealt{Wen.2020}), and it also exhibited weak short-term variability in some \xmm\ observations. \citet{Pasham.2019} reported a 131 s QPO in the X-ray light curve of \asa. However, the short-term variability of \asa\ is still too weak for the study of long-term evolution. Therefore, we only select the two longest \xmm\ observations of \asa\ to show the diversity of short-term X-ray variability of TDEs.

The second source is \igr\ (Ra=12$^{\rm h}$58$^{\rm m}$01.19$^{\rm s}$, Dec=+01$\degr$34$'$33.02$''$), which was firstly discovered by {\it INTEGRAL} during its observing period of 2-11, January, 2011 (\citealt{Walter.2011}). This source is associated with the central region of the Seyfert 2 galaxy NGC 4845, which is located at redshift $z=0.0041$. \xmm\ observed \igr\ in 2011-01-22 for a single exposure of 21 ks, which revealed one of the strongest short-term X-ray variability in all known TDEs (\citealt{Nikolajuk.2013}; see Figure~\ref{fig-lccompare}e). It was proposed that the central engine of \igr\ is similar to \sw16, but with a larger inclination angle of $\theta>30\degr$ relative to the observer (\citealt{Lei.2016}). However, \citet{Auchettl.2017} argued that \igr\ is more likely to be a changing-look AGN rather than a TDE. Although the origin of \igr\ is still under debate, its X-ray variability is rather rare for both AGN and TDEs. Therefore, we still adopt it for comparison.

\subsection{Data Reduction}
\subsubsection{\xmm\ EPIC}
The \xmm\ observations used in this paper are listed in Table~\ref{tab-obs}. The data were downloaded from the \xmm\ science archive\footnote{http://nxsa.esac.esa.int/nxsa-web/}, and reprocessed with the \xmm\ Science Analysis System (SAS, v19.0.0, \citealt{Gabriel.2004}) with the latest calibration files. In this work, we only use the data from the European Photon Imaging Camera (EPIC).

Firstly, the {\tt odfingest}, {\tt epchain} and {\tt emchain} tasks were used to reprocess the data and produce event files. Then we followed the SAS thread and used the {\tt evselect} task to extract background light curves with pattern zero in the high-energy range, i.e. 10-12 keV for EPIC-pn and $>$10 keV for MOS. These are used to identify good time intervals (GTIs) having low steady background count rates, so that the influence of soft proton flares is suppressed{\footnote{See \citet{Rosen.2016} for more details about different methods of flare filtering.}}. Then we performed further event filtering with {\sc FLAG=0}, patterns $\le 4$ for EPIC-pn and $\le 12$ for MOS.

The source extraction region was defined by a circular region of radius 50" centered at the source position. The background events were extracted from a nearby source-free region. Then the source and background spectra and light curves were extracted with the {\tt evselect} task in these regions. The pile-up effect was checked with the {\tt epatplot} task, and no significant pile-up was found. The response and auxiliary files were produced by the {\tt rmfgen} and {\tt arfgen} tasks. The {\tt grppha} task was used to regroup spectra with at least 25 counts per energy bin. Spectral fittings were performed with the {\sc xspec} software (v12.11.0m, \citealt{Arnaud.1996}), which is part of the {\sc heasoft} package (v6.28, \citealt{Blackburn.1995}).

The {\tt epiclccorr} task was ran to perform various corrections on the light curves{\footnote{https://heasarc.gsfc.nasa.gov/docs/xmm/sas/help/epiclccorr/index.html}}. An important point to note is that for all the light curves analyzed in this work, we did not apply the correction of point-spread-function (PSF) for the source flux. The reason is that a PSF correction will multiply a constant scaling factor of larger than 1 to the light curve, thereby increasing both the mean flux and fluctuation. Then the theoretical Poisson-noise power of 2/<rate> (e.g. \citealt{Uttley.2014}), where <rate> is the mean count rate, would be an under-estimate of the intrinsic Poisson power. As a result, the intrinsic variability power would be over-estimated.

Note that for the Obs-1 of \sw16, we made use of the spectra and light curves from MOS-1, because EPIC-pn and MOS-2 were both in the {\it Fast} mode. For all the rest of the observations, we used the data from EPIC-pn, as it provides data with the highest signal-to-noise ratio (S/N). Moreover, for the Obs-2 of \asa, the data in the first 10 ks and after 80 ks were severely affected by background flares, and so only the 70 ks data segment in the middle was used.

\subsubsection{\swift\ XRT}
The \swift/XRT data of \sw16\ provide a well-sampled long-term X-ray light curve, which also covers the observing window of the 13 \xmm\ observations. We use this long-term light curve to define the baseline flux evolution, and then identify the flux state of every \xmm\ observation. The data were downloaded from NASA's High Energy Astrophysics Science Archive Research Center (HEASARC)\footnote{https://heasarc.gsfc.nasa.gov/}.

The {\sc heasoft} package (v6.28) was used for the data reduction. Firstly, the {\tt xrtpipeline} task was used to reprocess the XRT data. The source extraction region was defined as a circular region of 30" radius, which was used in both the photon-counting (PC) mode and the window-timing (WT) mode, although in the WT mode only a single row of pixels was read out. The background events were extracted from nearby source-free regions. The event filtering and the extraction of spectra and light curves were all conducted with the {\tt xselect} task. The {\tt xrtexpomap} and {\tt xrtmkarf} tasks were used to produce the exposure map, response and auxiliary files. Similarly, the {\tt grppha} task was used to regroup spectra with at least 20 counts per energy bin. Since the purpose of using XRT data is to reveal the long-term flux variation, we used an absorbed power law model to fit the XRT spectrum in every observation, and measured the 0.3-10 keV flux from the best-fit model.

\begin{figure}
\centering
\includegraphics[trim=0.05in 0.1in 0in 0in, clip=1, scale=0.56]{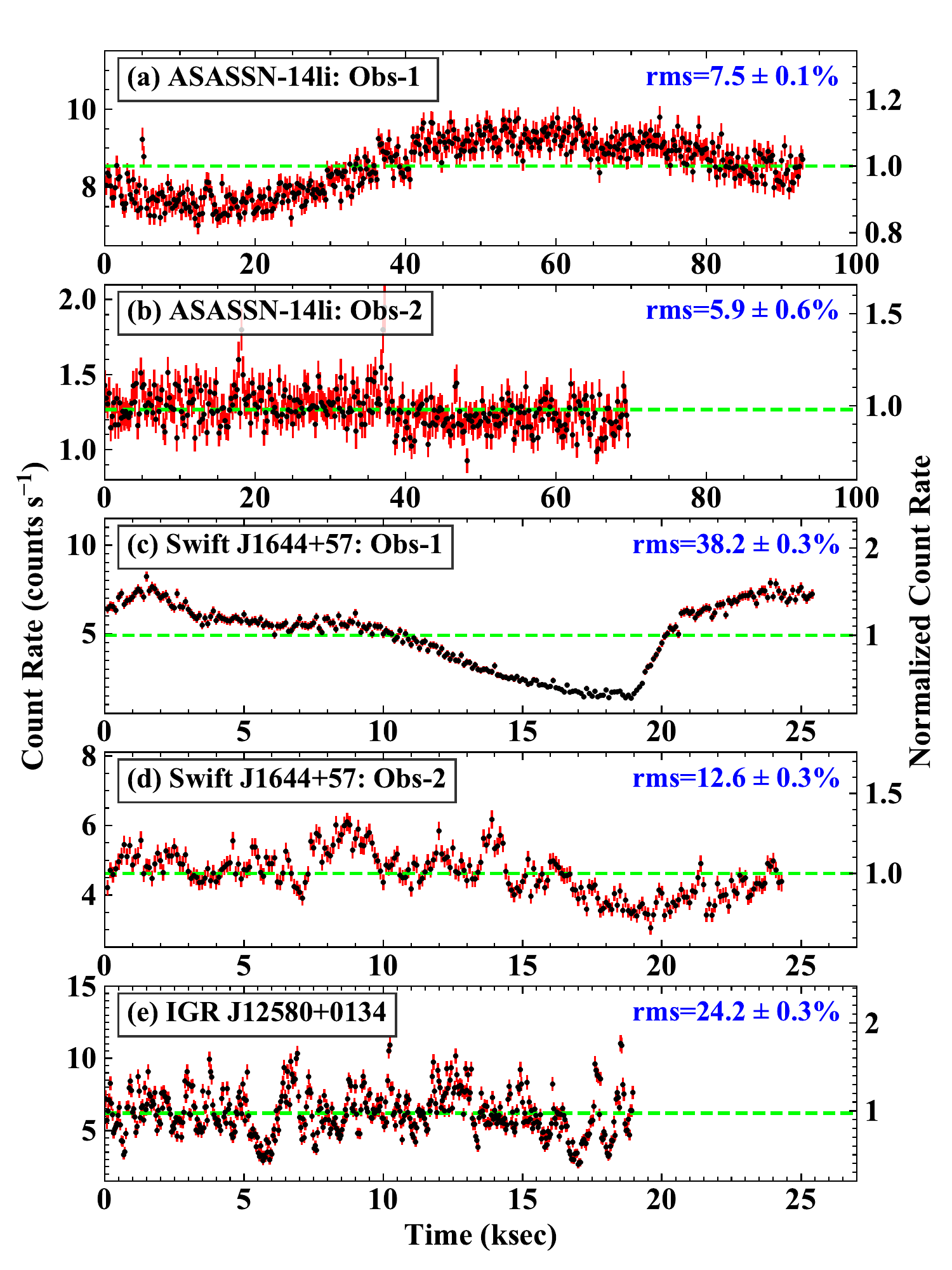}
\caption{Miscellaneous 0.3-10 keV short-term light curves of some TDEs observed by \xmm. The labelled rms value is the total fractional variability amplitude.}
\label{fig-lccompare}
\end{figure}

\begin{figure}
\centering
\includegraphics[trim=0.05in 0.1in 0in -0.2in, clip=1, scale=0.62]{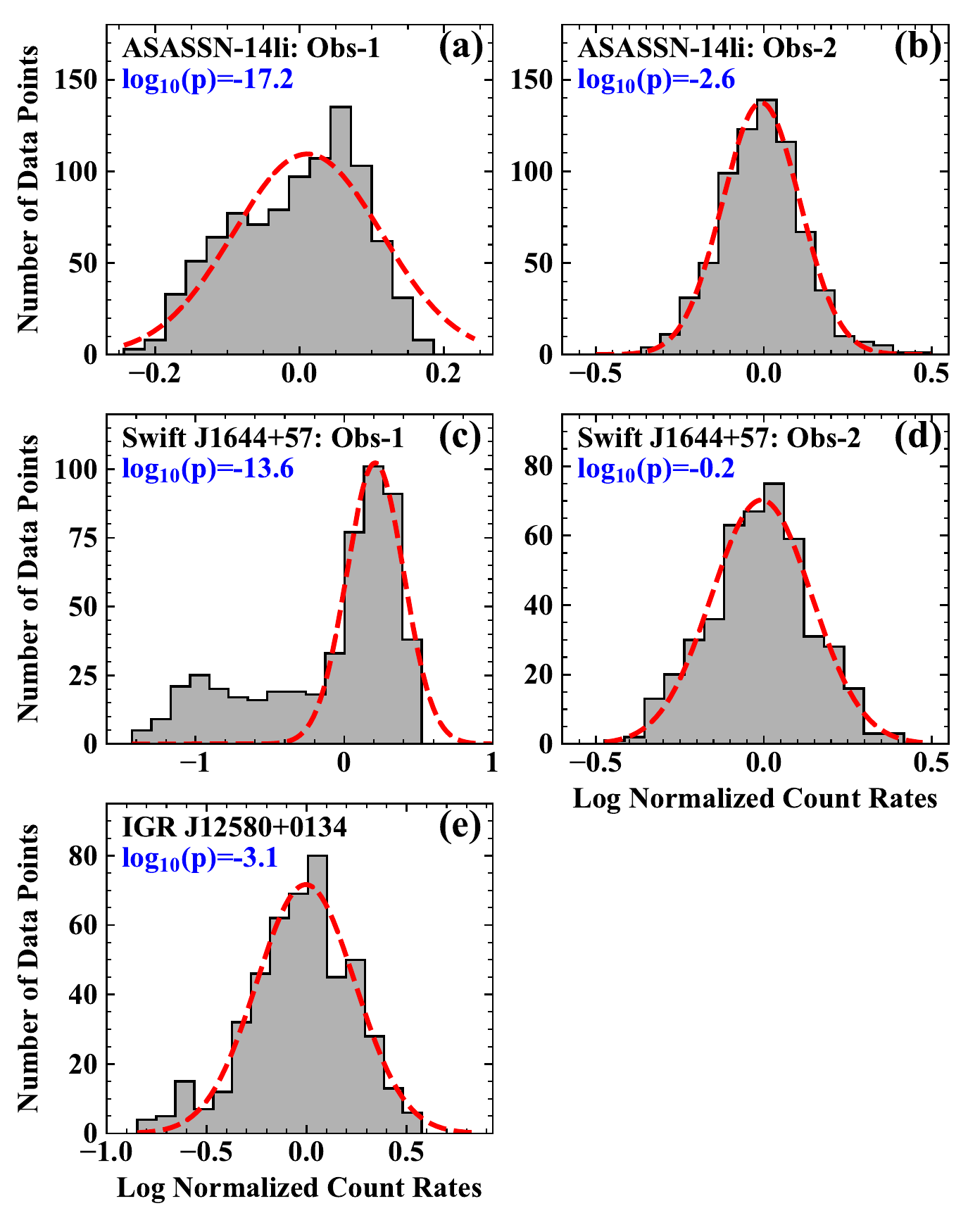}
\caption{The distribution of logarithmic normalized count rates of the TDE light curves in Fig.\ref{fig-lccompare}. The red dash line shows the best-fit normal distribution. The labelled $p$-value indicates the deviation of the observed flux distribution from the lognormal distribution.}
\label{fig-lograte}
\end{figure}

\section{The Diversity of TDE's Short-term X-ray Variability}
\label{sec-dvar}
Figure~\ref{fig-lccompare} shows the 0.3-10 keV light curves of the three TDEs. Firstly, it can be seen that different TDEs can exhibit very different short-term X-ray variability. Secondly, it is clear that the same TDE can also show very different short-term X-ray variability in different observations. These two results can be quantified by the fractional rms amplitude (e.g. \citealt{Edelson.2002}; \citealt{Arevalo.2008}), which are labelled in every panel of Figure~\ref{fig-lccompare}. The rms of \asa\ is 7.5\% in Obs-1 and 5.9\% in Obs-2. The rms of \sw16\ is 38.2\% in Obs-1, but decreases to 12.6\% in Obs-2. The large rms of \sw16\ in Obs-1 is mainly caused by the monochromatic flux decrease in the 10-19 ks segment, as well as the rapid flux increase right after 19 ks. The typical timescale of the short-term variability of \sw16\ is also longer in Obs-1 than in Obs-2. The rms of \igr\ is 24.2\%, and it shows the shortest variability timescales among the three TDEs. It is known for AGN that there exists a significant linear correlation between the X-ray rms and the black hole mass (e.g. \citealt{Ponti.2012}). If this relation also holds for TDEs, then \asa\ may have the highest black hole mass among the three TDEs, while \igr\ may have the lowest black hole mass.

The observed variation of variability patterns between different observations of the same TDE implies that the source may undergo different physical processes as it evolves. A simple test is to check the flux distribution of the light curve. It is known that the X-ray flux of AGN and XRB often exhibits a lognormal distribution (\citealt{Vaughan.2003a, Vaughan.2003b}; \citealt{Gaskell.2004}). This provides a fundamental evidence that the radiation processes in the X-ray corona should be multiplicative rather than additive (\citealt{Aitchison.1963}; \citealt{Uttley.2005}). Hence if the flux distribution changes significantly, it would be a strong indication that the underlying physical process has changed.

For the five light curves in Figure~\ref{fig-lccompare}, we derive their flux distributions and compare them with the lognormal distribution, as shown in Figure~\ref{fig-lograte}. The $p$-value is used to quantify the deviation of the observed flux distribution from lognormal. We find that the flux distribution changes from one observation to another for the same TDE. For \asa, the flux distribution in Obs-1 appears double-peaked rather than being a single lognormal distribution, which is confirmed by the small $p$-value of $10^{-17.2}$. However, it exhibits a much better lognormal distribution in Obs-2, with a $p$-value of $10^{-2.6}$. Similarly, \sw16\ shows a double-peaked flux distribution in Obs-1, but an almost perfect lognormal flux distribution in Obs-2. \igr\ also shows a good lognormal flux distribution.

Therefore, our investigation of flux distribution suggests that different physical processes are likely to cause the short-term variability of \asa\ and \sw16\ during different evolution stages.

\begin{table*}
\begin{minipage}{175mm}
  \centering
\caption{Spectral and variability properties of \sw16\ measured in individual \xmm\ observations.}
   \label{tab-pars}
\begin{tabular}{@{}lcccccccccc@{}}
\hline
\hline
Obs & $F_{\rm 0.3-10}$ & $\Gamma_{\rm 0.3-10}$ & $N_{\rm H}$ & $\chi^2$/dof & $r_{\rm flux}$ & $p$-value & $s_{\rm PSD}$ & Rms & $\gamma^2$ & $t_{\rm lag}$ \\
 & & & ($10^{22}$ cm$^{-2}$) & & & (log) & & (\%) & & (sec)\\
\hline
1$^{*}$ & 7.73$\pm$0.20  & 1.85$\pm$0.02& 1.05$\pm$0.02 & 551.0/452 & 0.45 & -15.12  & 2.81$\pm$0.47 & 22.4$\pm$8.6 & 0.99$\pm$0.00 & 52.1$\pm$12.2 \\
2 & 6.08$\pm$0.10  & 1.75$\pm$0.01 & 1.01$\pm$0.01 & 1507.6/1449 & 0.78 & -0.16  & 1.54$\pm$0.21 & 10.5$\pm$2.7 & 0.93$\pm$0.03 & -20.2$\pm$36.1 \\
3 & 5.02$\pm$0.12 & 1.56$\pm$0.02 & 1.06$\pm$0.02 & 1261.5/1253 & 1.06 & -3.37  & 1.43$\pm$0.17 & 11.2$\pm$2.4 & 0.96$\pm$0.02 & -0.5$\pm$24.9 \\
4$^{*}$ & 0.42$\pm$0.03 & 1.90$\pm$0.05 & 0.97$\pm$0.05 & 324.5/302 & 0.14 & -8.09 & 2.21$\pm$0.31 & 48.8$\pm$14.7 & 0.99$\pm$0.01 & 44.7$\pm$13.5  \\
5 & 1.43$\pm$0.04 & 1.56$\pm$0.02 & 1.09$\pm$0.03 & 994.0/999 & 0.65 & -4.13 & 1.55$\pm$0.24 & 17.3$\pm$4.2 & 0.94$\pm$0.02 & 20.7$\pm$32.2  \\
6$^{*}$ & 0.14$\pm$0.02 & 1.58$\pm$0.08 & 1.01$\pm$0.09 & 155.1/127 & 0.12 & -15.16 & 0.77$\pm$0.51 & 14.8$\pm$4.1 & 0.09$\pm$0.14 & 98.2$\pm$478.3  \\
7 & 1.16$\pm$0.04 & 1.41$\pm$0.03 & 1.16$\pm$0.03 & 965.7/911 & 1.18 & -0.92 & 1.60$\pm$0.26 & 18.5$\pm$4.6 & 0.95$\pm$0.02 & -15.2$\pm$26.9  \\
8 & 0.98$\pm$0.05 & 1.41$\pm$0.03 & 1.15$\pm$0.04 & 568.2/631 & 1.17 & -0.45 & 1.57$\pm$0.27 & 28.9$\pm$10.1 & 0.91$\pm$0.04 & -93.3$\pm$45.2  \\
9 & 1.05$\pm$0.04  & 1.34$\pm$0.02 & 1.14$\pm$0.03 & 916.5/921 & 1.57 & -3.10 & 1.12$\pm$0.19 & 12.8$\pm$2.6 & 0.83$\pm$0.06 & 70.9$\pm$55.7  \\
10$^{*}$ & 0.24$\pm$0.02  & 1.31$\pm$0.06 & 1.01$\pm$0.07 & 268.8/289 & 0.42 & -16.95 & 1.96$\pm$2.24 & 10.0$\pm$2.5 & 0.33$\pm$0.15 & -126.2$\pm$170.6  \\
11 & 0.41$\pm$0.02  & 1.39$\pm$0.04 & 1.12$\pm$0.05 & 451.5/430 & 0.81 & -5.77 & 1.31$\pm$0.35 & 15.8$\pm$4.3 & 0.59$\pm$0.12 & 49.3$\pm$97.1  \\
12$^{*}$ & 0.26$\pm$0.02  & 1.36$\pm$0.06 & 1.04$\pm$0.08 & 211.8/222 & 0.58 & -41.95 & 0.85$\pm$0.36 & 14.5$\pm$3.8 & 0.44$\pm$0.17 & 155.6$\pm$145.5  \\
13 & 0.58$\pm$0.04  & 1.32$\pm$0.05 & 1.04$\pm$0.06 & 321.9/331 & 1.46 & -9.13 & 1.36$\pm$0.26 & 18.5$\pm$4.4 & 0.72$\pm$0.09 & -57.5$\pm$73.2  \\
\hline
\end{tabular}
\end{minipage}
\\
\\
Notes. $F_{\rm 0.3-10}$ is absorbed flux in 0.3-10 keV in units of $10^{-11}$ erg cm$^{-2}$ s$^{-1}$. $\Gamma_{\rm 0.3-10}$ and $N_{\rm H}$ are the best-fit photon index and Hydrogen column density. $\chi^2$/dof is the corresponding chi square divided by the degrees of freedom. These are obtained by fitting an absorbed power law model to the 0.3-10 keV spectrum. $r_{\rm flux}$ is the ratio between the observed flux and the baseline flux trend in Figure~\ref{fig-evo-spec}a. $p$-value indicates the deviation of the observed flux distribution from the lognormal form. $s_{\rm PSD}$ is the slope of the PSD in 0.3-10 keV. The rms variability is measured in 2-10 keV in the frequency range of $(0.4-5)\times10^{-4}$ Hz. $\gamma^{2}$ and $t_{\rm lag}$ are the coherence and time lag measured between 0.3-1 keV and 2-10 keV in the same frequency range mentioned above. The $*$ symbol indicates that the observation caught \sw16\ in the dipping state.
\end{table*}

\section{Long-term Evolution of X-ray Spectral-timing Properties}
\label{sec-svar}
Because of the availability of the large amount of high-quality X-ray data, \sw16\ offers a good opportunity for us to investigate how various short-term X-ray spectral-timing properties evolve with the long-term TDE phase.

\subsection{X-ray Properties Investigated}
\label{sec-xray-prop}
It has been reported that the time-averaged X-ray spectrum of \sw16\ in 0.3-10 keV is consistent with an absorbed power law (\citealt{Burrows.2011}; \citealt{Saxton.2012}). Therefore, we use an absorbed power law to model the X-ray spectra observed in each \xmm\ and \swift\ observation, and check their long-term evolution. The neutral absorption is modelled with the {\sc tbabs} model in {\sc xspec} with the abundances of \citet{Wilms.2000} and cross-sections of \citet{Verner.1996}. The spectral properties include the observed flux in 0.3-10 keV, the equivalent hydrogen column density $N_{\rm H}$, and the power law photon index.

Then we select some parameters to describe the short-term X-ray variability. A basic property of an X-ray light curve is the flux distribution, which is mentioned in the previous section. Likewise, we use $p$-value to quantify the difference of the flux distribution from the lognormal form, and explore its long-term evolution.

Another useful method is the power spectral density (PSD). The X-ray PSD of an AGN generally consists of Poisson noise and red noise. Other features such as PSD breaks and QPO signals also emerge sometimes (e.g. \citealt{McHardy.2006, McHardy.2007}; \citealt{Gierlinski.2008}; \citealt{Jin.2020}). These components can also be found in TDEs' PSDs (e.g. \citealt{Reis.2012}; \citealt{Pasham.2019}). We use a simple PSD model consisting of two components. The first component is a free constant to account for Poisson noise. The second component is a single power law of the following form to account for the red-noise continuum:

\begin{equation}
P(f) = C_{\rm 0}\cdot f^{-s_{\rm PSD}}
\end{equation}
where $P$ is the power density, $C_{\rm 0}$ is the free normalization, $f$ is the frequency and $s_{\rm PSD}$ is the slope. We also adopt the Belloni-Hasinger normalization (\citealt{Belloni.1990}) for the PSD, so that one can integrate the PSD within a specific frequency range to derive the fractional rms amplitude. We follow the Bayesian approach described in \citet{Vaughan.2010} to perform the PSD analysis and measure model parameters.

The frequency-differentiated coherence and time lags measured between different energy bands can provide valuable information about the X-ray mechanism (e.g. \citealt{Fabian.2009, Fabian.2013}; \citealt{Uttley.2014}; \citealt{Kara.2016}; \citealt{Jin.2017a, Jin.2020, Jin.2021}). It must be noted that a time lag measurement is meaningful and robust only if it is associated with a high coherence. We calculate these two properties in individual \xmm\ observations. All the results are reported in Table~\ref{tab-pars} for every \xmm\ observations.

\begin{figure*}[ht]
\centering
\includegraphics[trim=0.1in 0.0in 0in -0.2in, clip=1, scale=0.65]{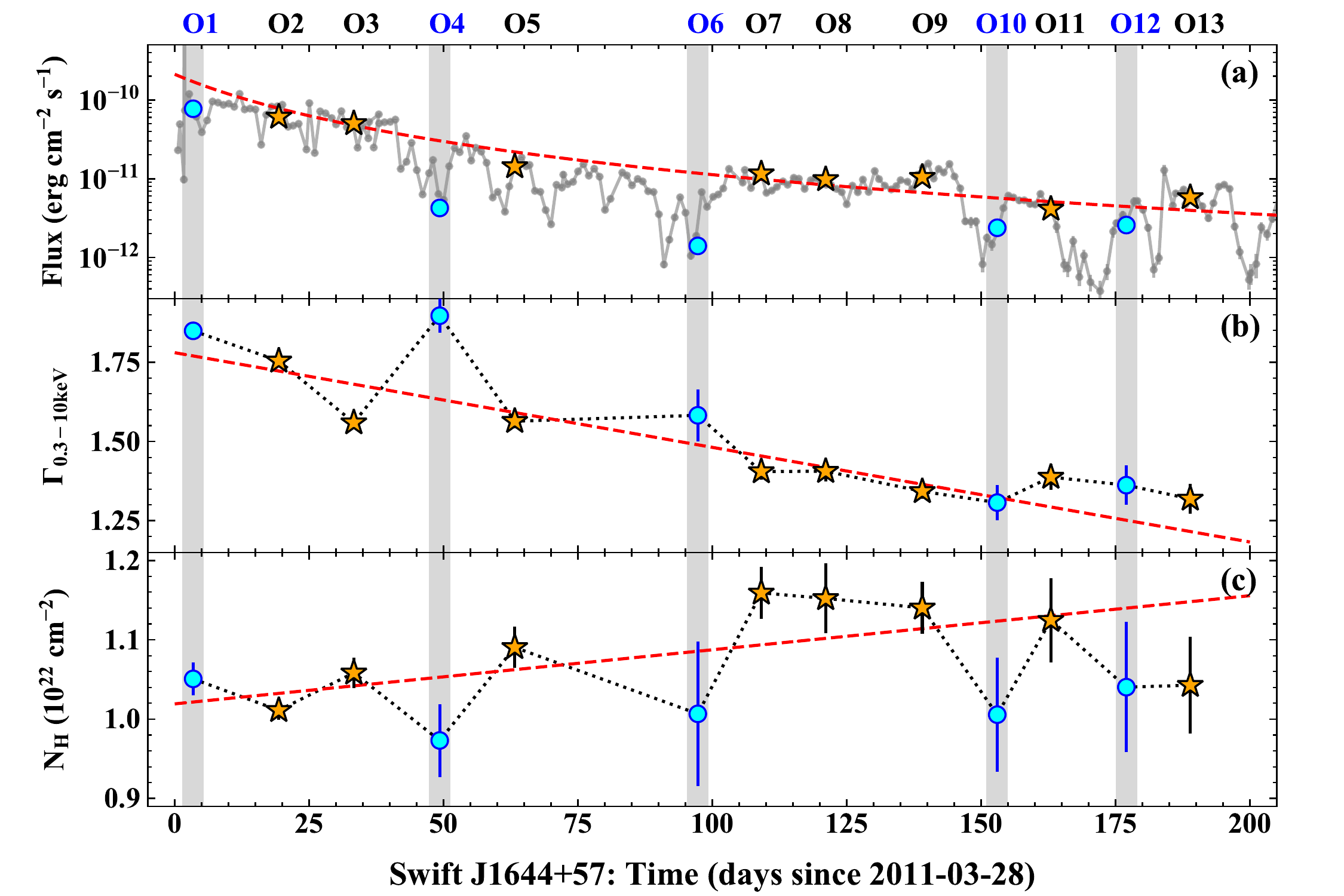}
\caption{The evolution of the flux and spectral parameters of \sw16. In panel-a, the gray points are from \swift/XRT. The cyan circles and orange stars are from \xmm\ observations inside and outside dips. The red dash line shows the baseline flux decay with a power law index of -5/3. Panels b and c present the evolution of the best-fit power law photon index in 0.3-10 keV and $N_{\rm H}$. The red dash lines in these two panels are the results of a linear fit to all the data points, which show the long-term evolution of the two parameters .}
\label{fig-evo-spec}
\end{figure*}

\begin{figure*}[ht]
\centering
\includegraphics[trim=0.1in 0.0in 0in -0.2in, clip=1, scale=0.65]{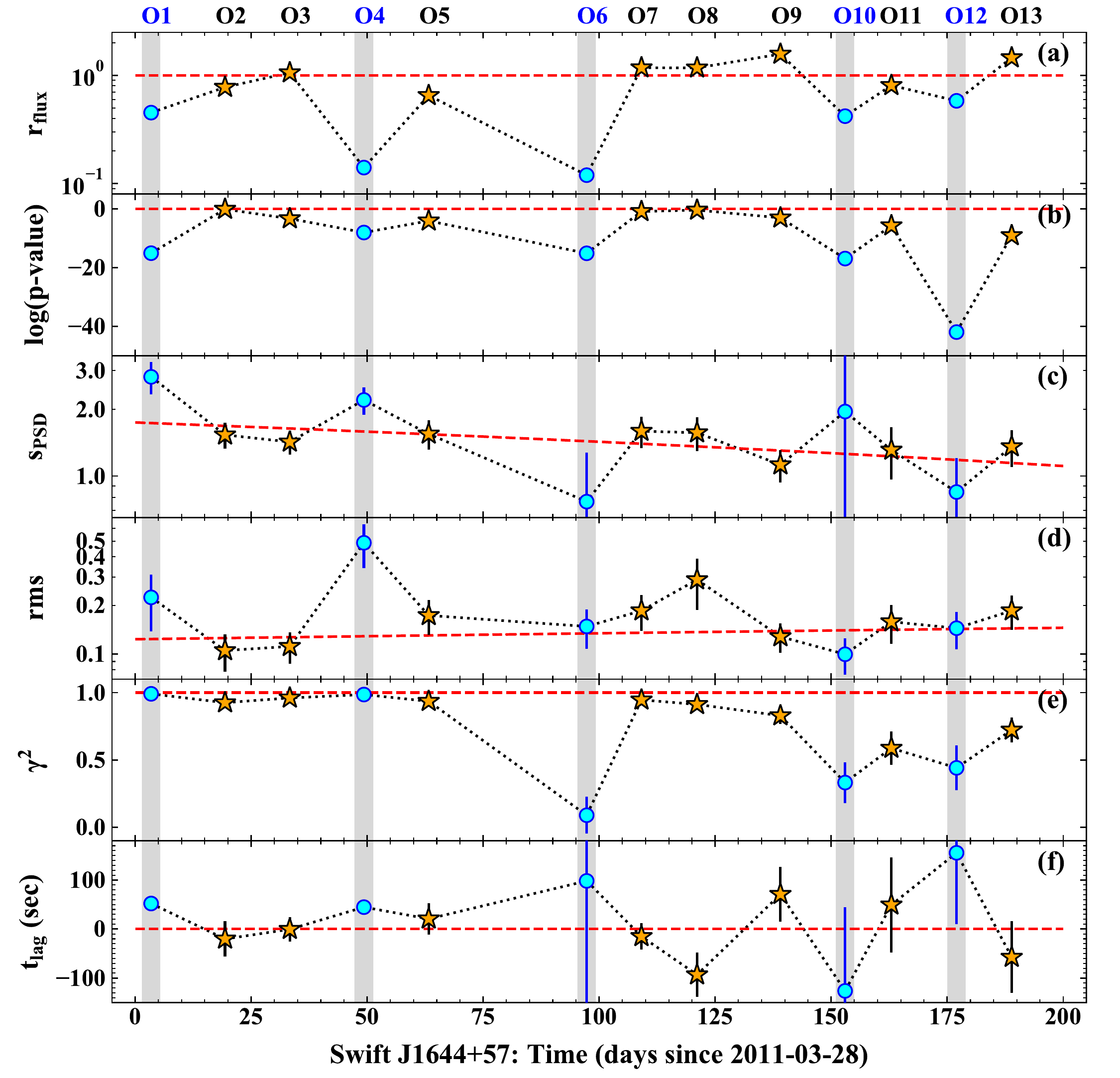}
\caption{The evolution of the short-term X-ray variability properties of \sw16. The values are listed in Table~\ref{tab-pars}. The cyan circles and orange stars are from \xmm\ observations inside and outside the dipping state. Panel-a shows the evolution of $r_{\rm flux}$, which is the flux relative to the baseline flux. Panel-b shows the evolution of $p$-value which indicates the deviation of the observed 2-10 keV flux distribution from the lognormal distribution. Panel-c shows the evolution of the PSD slope ($s_{\rm PSD}$) in 2-10 keV. Panel-d shows the evolution of the 2-10 keV fractional rms in the frequency range of $(0.4-5)\times10^{-4}$ Hz. In each of these two panels, the red dash line is the result of a linear fit to all the data points, which shows the trend of long-term evolution. Panels e and f present the evolution of the coherence ($\gamma^2$) and time lag ($t_{\rm lag}$) measured between 0.3-1 keV and 2-10 keV for the same frequency range mentioned above. The red dash lines in these two panels are for $\gamma^2$=1 and $t_{\rm lag}=0$.}
\label{fig-evo-var}
\end{figure*}

\subsection{X-ray Spectral Evolution}
Firstly, we check the long-term spectral evolution of \sw16. This was studied in great detail before, and our results are consistent with previous works (\citealt{Burrows.2011}; \citealt{Saxton.2012}; \citealt{Mangano.2016}). Figure~\ref{fig-evo-spec}a shows the long-term evolution of the flux in 0.3-10 keV, including both \xmm\ and \swift\ observations. The gray points are observed by \swift/XRT. The orange and cyan points are observed by \xmm/EPIC. This light curve shows that there is a long-term flux decay. The red dash line is a power law, which represents the baseline flux decay of \sw16. The slope of this power law is -5/3, which follows the theoretical prediction that the X-ray luminosity should follow the return rate of stellar debris (e.g. \citealt{Tchekhovskoy.2014}; also see \citealt{Saxton.2020} and references therein). There are also some short-term dipping periods when the source flux fell below the baseline flux decay significantly.

Then we divide the 13 \xmm\ data points into two classes, according to the observed flux relative to the baseline flux (i.e. $r_{\rm flux}$ in Table~\ref{tab-pars}). For the five observations with $r_{\rm flux} < 0.6$, we refer to them as dipping-state observations. The other eight observations are called normal-state observations.

Figures~\ref{fig-evo-spec}b and \ref{fig-evo-spec}c present the long-term evolution of the best-fit photon index and $N_{\rm H}$ in 0.3-10 keV. The red dash line is a linear fit to all the data points, showing the trend of the long-term evolution. For the photon index evolution, we find a linear slope of -0.0029 $\pm$ 0.0004. When we use a free constant to fit the data (i.e. the slope is fixed to zero), the $\chi^2$ increases by 522, and so the observed trend is highly significant. It shows that the spectrum becomes harder as the baseline flux declines, but it gets softer during the dipping state. This is consistent with the results reported by previous studies (\citealt{Saxton.2012}; \citealt{Mangano.2016}), and can be explained in terms of the variation of a relativistic jet (\citealt{Saxton.2012}; \citealt{Lei.2013}; \citealt{Tchekhovskoy.2014}). The linear fit to the $N_{\rm H}$ data has a slope of 0.00068 $\pm$ 0.00023. When a free constant is used to fit the data, the $\chi^2$ increases by 19, and so the observed trend of increasing $N_{\rm H}$ is significant. However, it must also be noted that $N_{\rm H}$ is directly degenerate with the photon index, thus the observed $N_{\rm H}$ evolution is model dependent.

\subsection{X-ray Variability Evolution}
Then we present the long-term evolution of various short-term X-ray variability properties. The corresponding values are all listed in Table~\ref{tab-pars}. We also plot the spectra, light curves, flux distributions, PSDs, lag spectra and coherence spectra for individual \xmm\ observations in Appendix~\ref{app-sec-plots}.

Figure~\ref{fig-evo-var}a shows the evolution of $r_{\rm flux}$ for all the \xmm\ observations. The error bars are too small to be visible. These are inherited from the flux errors given in Table~\ref{tab-pars}. This panel shows clearly that Obs-1, 4, 6, 10, 12 are in the dipping state, while the other observations are in the normal state.

Figure~\ref{fig-evo-var}b shows how the $p$-value of the short-term flux distribution evolves. We find that \sw16\ exhibits a good lognormal flux distribution during the normal state, as indicated by the orange stars. In comparison, the $p$-value decreases significantly during the dipping state, suggesting that the flux distribution deviates significantly from the lognormal form. These results indicate that the X-ray mechanism in the dipping state should be different from the normal state. This is qualitatively consistent with the jet-precession model, where the X-ray emission in the normal state is dominated by the extended part of the jet cone, while in the dipping state the emission from the jet base becomes more important (\citealt{Saxton.2012}; \citealt{Lei.2013}; \citealt{Tchekhovskoy.2014}).

Figure~\ref{fig-evo-var}c shows the long-term evolution of the PSD slope in 2-10 keV. Firstly, we find a weak trend of decreasing PSD slope, as indicated by the red dash line which is a linear fit to the data. The slope of this line is -0.0032 $\pm$ 0.0016. When a free constant is used to fit the data, the best-fit PSD slope is 1.46 $\pm$ 0.10, and the $\chi^2$ increases by 6.3, which is equivalent to a significance of 2.5 $\sigma$. This weak trend indicates that the high-frequency variability becomes more important as \sw16\ evolves. If we consider the variability timescale as an indicator of the size of the X-ray emission region, then our result suggests that the X-ray emission region should be contracting. Secondly, our study shows that the PSD slope observed in Obs-1 and 4, which are the first two \xmm\ observations in the dipping sate, exhibits a significant increase comparing to adjacent observations in the normal state. This result can be confirmed by the light curves shown in Appendix~\ref{app-sec-plots}, which shows that the variability patterns observed in Obs-1 and 4 are significantly different from the other observations. Specifically, the low-frequency variability in these two observations appears much stronger, which leads to their steeper PSDs.

Figure~\ref{fig-evo-var}d shows the long-term evolution of the fractional rms variability in 2-10 keV in the frequency range of $(0.4-5)\times10^{-4}$ Hz. The lower frequency limit is set by the length of each light curve, which is $\sim0.4\times10^{-4}$ Hz for the longest observation. The corresponding variability timescale is from 2 ks to $\sim$28 ks. Similarly, we find that Obs-1 and 4 show significantly larger rms values than their adjacent observations. This is also caused by the extra low-frequency variability in these two observations. Among the remaining 11 observations, only Obs-8 has a relatively large rms of 0.29 $\pm$ 0.10. The other observations all have rms values between 0.1 and 0.2, including the later three dipping-state observations. A linear fit to all the 13 rms points has a slope of 0.0001 $\pm$ 0.0002, as indicated by the red dash line. When we use a free constant to fit the data, the $\chi^2$ only increases by 0.42, which is equivalent to a significance of 0.65 $\sigma$. Therefore, there is no significant trend of baseline rms evolution.

Figure~\ref{fig-evo-var}e shows the long-term evolution of the coherence between light curves in 0.3-1 keV and 2-10 keV. The red dash line is for $\gamma^2=1$, which indicates the position of a perfect coherence. The inspected frequency range is also from $\sim3.6\times10^{-5}$ Hz to $5\times10^{-4}$ Hz. Firstly, high coherences are found during the first five observations independent of the flux states. Secondly, there is a weak decline of coherence during the normal state, but a significant drop of coherence during the dipping state. A simple explanation for the variation of coherence is the decrease of flux, because when the flux is lower the light curve is more white-noise dominated, and so the inter-band coherence is also lower. Considering the fact that the first five observations all exhibit high coherences, it is likely that the coherence does not have an intrinsic dependence on the flux state.

Figure~\ref{fig-evo-var}f shows the long-term evolution of the time lag between 0.3-1 keV and 2-10 keV for the same frequency range as the coherence. Significant time lags are observed in Obs-1 and 4, but not in any other observations. The robustness of these lags are ensured by their high coherences. The time lag is 52.1 $\pm$ 12.2 s in Obs-1, and 44.7 $\pm$ 13.5 s in Obs-4, both indicating that the soft X-ray variation lags behind the hard X-ray.

To explore these lags in more detail, the lag and coherence spectra are produced for both Obs-1 and 4, which are presented in Figure~\ref{fig-cohlag} (see Appendix~\ref{app-sec-plots} for the results of all observations). The reference band is 2-10 keV. In both observations, high coherences are found across the 0.3-10 keV band. The slight decrease of coherence above 6 keV is likely due to the decrease of count rate. The coherence spectra ensure that all the lag measurements are robust. The lag spectrum of Obs-1 shows that there is no significant non-zero lag above 3 keV, but the lag increases significantly from 3 keV to 0.3 keV. The total lag between 0.3-1 keV and 3-6 keV is 76.2 $\pm$ 27.4 s. The lag spectrum in Obs-4 shows a similar shape and amplitude, except for larger statistical uncertainties due to the lower count rate than in Obs-1. We discuss the physical implication of these lags in the following section.

\begin{figure}
\centering
\includegraphics[trim=0.1in 0.0in 0in -0.1in, clip=1, scale=0.65]{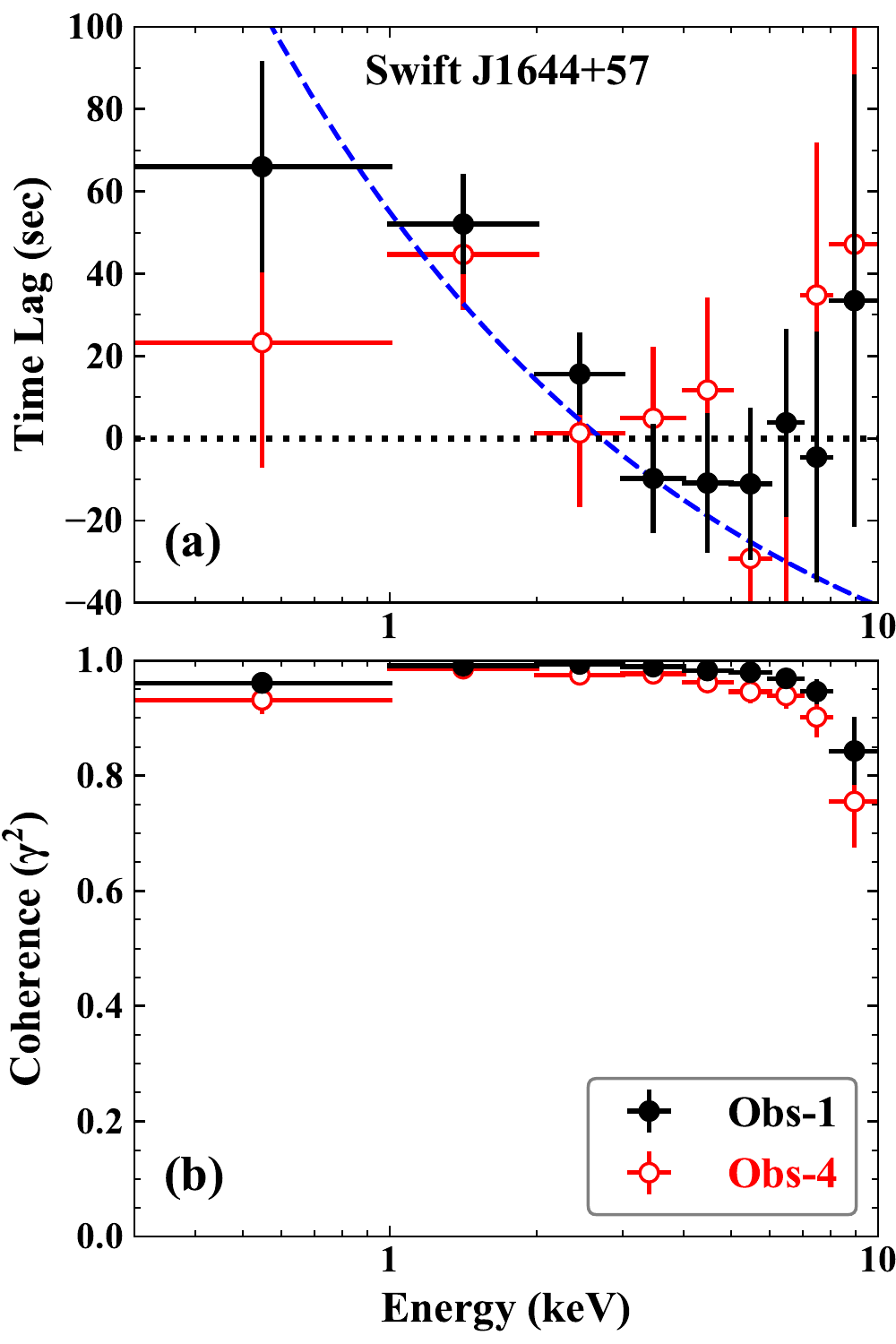}
\caption{The lag and coherence spectra of \sw16\ observed in Obs-1 (black) and Obs-4 (red). The reference band is 2-10 keV. The inspected frequency range is $(0.4-5)\times10^{-4}$ Hz. The blue dash line shows the $t_{\rm lag} \propto E^{-1/2}$ relation with an arbitrary normalization for the synchrotron cooling model.}
\label{fig-cohlag}
\end{figure}

\section{Discussion}
\subsection{Constraining the X-ray Mechanism of \sw16}
The results of this work provide new evidence to constrain the X-ray mechanism of \sw16. The most important result is the significant difference of short-term X-ray variability between the normal and dipping states, as shown by the long-term evolution of various spectral timing properties.

Firstly, we find that the short-term X-ray flux of \sw16\ exhibits the form of lognormal distribution during the normal state, similar to the results found in AGN and XRBs (e.g. \citealt{Vaughan.2003a, Vaughan.2003b}; \citealt{Gaskell.2004}). This suggests that the physical processes producing these X-rays should be multiplicative (\citealt{Uttley.2005}). However, the flux distribution in the dipping state deviates from lognormal significantly, indicating extra/different physical processes in this state. This is qualitatively consistent with the jet-precession model (e.g. \citealt{Saxton.2012}), which proposes different origins for the X-ray emission in the normal and dipping states. However, it is still not clear why the X-ray emission in the dipping state does not have a lognormal distribution.

Secondly, we find that during the first two \xmm\ observations in the dipping state, \sw16\ exhibits significant low-frequency variability, which causes steeper PSDs and higher rms values. This does not happen in later observations regardless of flux states. Moreover, we find significant soft X-ray lags with high coherences in these two observations. With the reference band being 2-10 keV, the lag is $\sim$ 50 s for 0.3-1 keV, and it increases monochromatically from 0 s at 3 keV to $\sim$ 70 s at 0.3 keV. These new properties should also be explained by any models proposed for \sw16\ in the future.

The lag spectrum of \sw16\ also reminds us of the X-ray emission from some jetted AGN. For example, Mrk 421 is a BL Lac object whose X-ray emission is dominated by a relativistic jet along the line-of-sight, similar to the situation of \sw16. \citet{Takahashi.1996} reported an increasing soft X-ray lag below 2 keV for Mrk 421. One interpretation of this soft X-ray lag is the synchrotron cooling of relativistic electrons (e.g. \citealt{Tashiro.1995}). However, the lag in Mrk 421 is $\sim$ 5000 s, which is two orders of magnitude higher than in \sw16. Then the magnetic field in the jet of \sw16\ would be more than one order of magnitude higher for the same mechanism. We also notice other differences between the two sources. The lag spectrum of Mrk 421 follows the typical $t_{\rm lag} \propto E^{-1/2}$ relation (\citealt{Takahashi.1996}), but the lag in \sw16\ does not change significantly above 3 keV (see Figure~\ref{fig-cohlag}a). Besides, no lag is detected in any other \xmm\ observations in the later TDE phase regardless of flux states. Therefore, it is still too early to claim that the soft X-ray lag in \sw16\ has a similar mechanism to Mrk 421.

Instead, the shape of the lag spectrum of \sw16\ implies that there might exist a hard X-ray component dominating the energy band above 3 keV. Meanwhile, a separate soft X-ray component becomes significant during the dipping state, whose variability lags behind the hard X-ray component. This is also consistent with the analysis of spectral energy distribution in \citet{Burrows.2011}, where they speculated that the kink in the hard X-ray spectrum during the dipping state could be due to another component. This two-component scenario provides a straightforward explanation for the results of this work, as the stronger soft X-ray component can be responsible for causing various differences between the two states. However, the origins of these two components are still unclear, and it also needs to explain why the soft X-ray lag disappears during the later TDE phase.

\subsection{Estimating the Black Hole Mass of \sw16}
The black hole mass of an AGN can be estimated by various methods, such as reverberation mapping, single-epoch broad H$\beta$ line-width, correlations between the black hole mass and host galaxy properties or X-ray variability properties (e.g. rms, PSD break, QPO). However, it is much more difficult to estimate the black hole mass of a TDE. Firstly, optical broad lines of TDEs are not strongly correlated with their black hole masses (e.g. \citealt{Saxton.2018}). Indeed, it has been proposed that the width of optical broad lines of TDEs can be severely affected by electron scattering in a high-density gas (\citealt{Roth.2018}). Secondly, correlations based on the host galaxy properties can only provide order-of-magnitude estimates for the mass of the black hole in the center (e.g. \citealt{Kormendy.2013}). Thirdly, correlations based on the X-ray variability properties are not well established for TDEs, although such correlations might still hold if TDEs also have X-ray corona whose properties are similar to those in AGN.

For \sw16, a range of black hole masses have been reported (see Table~\ref{tab-mass}). For example, using the QPO signal, \citet{Reis.2012} estimated a black hole mass of $(0.45-5)\times10^{6}M_{\odot}$ (also see \citealt{Abramowicz.2012}). \citet{Miller.2011} used the radio and X-ray luminosities to derive a black hole mass of $3.2\times10^{5}M_{\odot}$. A mass of $\lesssim10^{5}M_{\odot}$ was reported by \citet{Krolik.2011} assuming the white-dwarf disruption model. In comparison, using the smallest X-ray variability timescale of $\sim$100 s as a rough estimate of the light crossing time of Schwarzschild radius, \citet{Burrows.2011} estimated a mass of $\sim7\times10^{6}M_{\odot}$. Using the host galaxy bulge luminosity, \citet{Burrows.2011} estimated the upper limit of the mass to be $\sim2\times10^{7}~M_{\odot}$. \citet{Yoon.2015} measured the stellar mass of the host galaxy's bulge, and reported a black hole mass of $10^{6.7\pm0.4}M_{\odot}$. Therefore, the previous mass estimates of \sw16\ give a range of values spanning more than one order of magnitude.

The 2-10 keV rms variability can also be used to infer the black hole mass (e.g. \citealt{Ponti.2012}). As shown in Table~\ref{tab-pars}, the range of rms of \sw16\ in the normal state is $0.10-0.30$, which corresponds to a mass range of $(1.3-7.9)\times10^{6}M_{\odot}$. The range of rms in the dipping state is $0.10-0.50$, thus the estimated mass range is $(0.6-7.9)\times10^{6}M_{\odot}$. These two mass ranges are both consistent with previous results. Since the jet emission may dominate the normal state, it may introduce extra short-term variability if the jet has substructures. Then the black hole mass can be under-estimated. However, it was also reported that for radio-loud, $\gamma$-ray detected NLS1s whose X-ray emission may also be dominated by powerful jets, X-ray variability can still be used to infer the black hole mass with a reasonable accuracy (e.g. \citealt{Pan.2018}). Therefore, the black hole mass estimates based on the X-ray rms of \sw16\ may still be reasonably good.

It is also interesting to compare \sw16\ with \igr, because the mass of \igr\ was reported to be as low as $\sim3\times10^{5}M_{\odot}$ (\citealt{Nikolajuk.2013}). Considering that the variability timescale of \igr\ is shorter than \sw16\ (see Figure~\ref{fig-lccompare}), it can be inferred that the mass of \sw16\ may be larger by a factor of few, so that a mass of $\sim10^{6}M_{\odot}$ is also preferred for \sw16.

However, unlike the X-ray emission of AGN, the X-ray variability properties of TDEs can evolve significantly as their fluxes decay, like \sw16\ which we studied in this work. Therefore, it is necessary to know at which stage of a TDE's evolution can its X-ray rms be the best indicator of the black hole mass. The situation of \sw16\ is further complicated by its jetted nature. For normal TDEs, it is probably the best to use the rms when the hard X-ray corona above 2 keV is formed (e.g. \citealt{Wevers.2021}). It is probably also the best to wait for the corona to be stablized, so that its properties become stationary, such as forming a typical lognormal flux distribution. Other possibilities of mass estimates include using some PSD features such as the high-frequency break and QPO to infer the black hole mass. However, a lot more observations are required in the future in order to investigate if these X-ray properties of TDEs also show long-term evolution.

\begin{table}
\centering
   \caption{Black Hole Mass Estimates for \sw16.}
    \label{tab-mass}
\begin{tabular}{lccc}
\hline
\hline
 Method & Value & $M_{\rm BH}$ & References \\
&  & ($10^{6}M_{\odot}$) & \\
\hline
X-ray QPO & $\sim$200 s & $0.1-5$ & 1, 2 \\
$L_{\rm radio}$ \& $L_{\rm X-ray}$ & -- &  $\sim0.32$ & 3 \\
WD Model & -- & $\lesssim0.1$ & 4 \\
$t_{\rm min}$ & $\sim$100 s & $\sim7$ & 5 \\
$L_{\rm B,bulge}$ & $1.6\times10^{9}L_{\odot,B}$ & $\lesssim20$ & 5 \\
$M_{\rm \star,bulge}$ & $1.4\times10^{9} M_{\odot}$ & $2-13$ & 6 \\
\hline
rms-normal & $0.10-0.30$ & $1.3-7.9$ & this work \\
rms-dipping & $0.10-0.50$ & $0.6-7.9$ & this work \\
\hline
\end{tabular}
\\
\smallskip
Notes. $L_{\rm radio}$ and $L_{\rm X-ray}$ are the radio and X-ray luminosities. {\it WD model} means the white-dwarf disruption model. $t_{\rm min}$ is the minimal variability timescale. $L_{\rm B,bulge}$ is the B-band luminosity of the galactic bulge. $M_{\rm \star,bulge}$ is the stellar mass of the bulge. {\it Rms-normal} is the 2-10 keV rms variability observed in the normal state, while {\it rms-dipping} is for the dipping state. References: (1) \citealt{Reis.2012}; (2) \citealt{Abramowicz.2012}; (3) \citealt{Miller.2011}; (4) \citealt{Krolik.2011}; (5) \citealt{Burrows.2011}; (6) \citealt{Yoon.2015}.
\end{table}

\subsection{Prospects of Monitoring TDE's Short-term X-ray Variability}
We have taken \sw16\ as an example to demonstrate the value of studying the long-term evolution of the short-term X-ray variability. \sw16\ is a special TDE because its X-ray emission is likely jet-dominated. As for normal TDEs, similar studies can be conducted if they are also monitored by a series of deep observations. However, it is worth noting that for some TDEs the short-term X-ray variability can be weak in some observations. This can happen if the X-ray emission is from the accretion disc itself rather than corona (e.g. \citealt{Wen.2020}), because the disc emission is often more stable than the corona, similar to the results found in super-Eddington NLS1s (e.g. \citealt{Jin.2017a, Jin.2017b}). However, as the hard X-ray corona emerges, the X-ray variability may increase significantly. This phenomenon has been observed in a rapid state transition of the TDE AT2018fyk (\citealt{Wevers.2021}).

In the coming era of time-domain astronomy, the number of newly discovered TDEs will increase significantly. For example, the {\it Einstein Probe} mission ({\it EP}: \citealt{Yuan.2018}), which is equipped with 12 lobster-eye X-ray telescopes with a total field-of-view of 3600 deg$^2$, may discover at least tens of X-ray TDEs per year (\citealt{Liu.2018}). It is not enough to monitor these TDEs with X-ray satellites in the low-earth orbit, such as \swift, \nustar\ (\citealt{Harrison.2013}), {\it NICER} (\citealt{Gendreau.2012, Gendreau.2016}) and {\it EP}. This is because these instruments cannot provide continuous long light curves for the study of short-term variability. A series of deep observations with X-ray telescopes of larger effective areas and longer orbital periods are needed, such as \xmm, {\it eROSITA} (\citealt{Merloni.2012}; \citealt{Predehl.2016}) and {\it Athena} (\citealt{Nandra.2013}).

With reference to the spectral-timing studies of super-Eddington NLS1s (e.g. \citealt{Jin.2013}), a continuous exposure of $\sim$100 ks per observation would be valuable to constrain most of the short-term variability properties, such as various PSD features (e.g. QPO, breaks), frequency-resolved rms, coherence and time lags. Then a series of similar observations are needed to follow the evolution of these properties. These observations, complemented by multi-wavelength monitoring campaigns of many short exposures, can provide a full range of detailed information about the dynamical evolution of TDEs, such as the formation and evolution of the accretion disc, corona, jet, inflow and outflow. It may also be possible to identify special TDE phases, when some X-ray variability properties (e.g. rms) can be used to make accurate estimates of the black hole mass. However, this strategy also heralds a significant increase of demands for long observations with various X-ray telescopes.

\section{Conclusions}
In this work, we use a set of high-quality data from \xmm\ and \swift\ to perform a case study on the well-observed jetted TDE \sw16, with special focus on its properties of short-term X-ray variability and their long-term evolutions. The main results are summarized below.

\begin{itemize}
\itemsep0em
\item We show that TDEs can exhibit diverse short-term X-ray variability patterns, which contain abundant information about the X-ray mechanism and physical properties.
\item We find that the short-term X-ray flux of \sw16\ in the normal state shows the form of lognormal distribution, but deviates from this form significantly in the dipping state.
\item We find that during the first two \xmm\ observations in the dipping-state (i.e. Obs-1 and 4), \sw16\ exhibited different low-frequency variability patterns, which lead to much steeper PSDs and larger fractional rms amplitudes.
\item Significant soft X-ray lags are detected with high coherences in both Obs-1 and 4, which are $\sim$ 50 s between 0.3-1 keV and 2-10 keV. But no significant lag is detected in later \xmm\ observations regardless of flux states.
\item We identify a potential long-term trend of PSD flattening, implying the contraction of the X-ray emission region.
\item The 2-10 keV rms for the frequency range of $(\sim0.36-5)\times10^{-4}$ Hz is found to be $0.10-0.30$ in the normal state and $0.10-0.50$ in the dipping state.
\item We use the 2-10 keV rms to estimate the black hole mass of \sw16, which is found to be $(0.6-7.9)\times10^{6}M_{\odot}$, consistent with previous results based on other methods. However, this rms method is severely affected by the evolution of short-term X-ray variability as the TDE evolves.
\end{itemize}

Our results add new constraints on the X-ray mechanism of \sw16. Our study also demonstrates the great potential of conducting similar studies for new TDEs. We suggest that while a long-term X-ray monitoring with a series of short exposures is necessary to follow TDE's evolution, it is also valuable to perform a series of deeper X-ray observations spreading over the TDE's lifetime, with each observation having a continuous exposure of a few tens to hundreds of ksec. Therefore, we foresee a significant increase of demands for deep observations using X-ray telescopes with large effective areas and long orbital periods.

\acknowledgments
We thank the anonymous referee for providing a thorough review and valuable comments. CJ thanks Wenda Zhang, Erlin Qiao, Haiwu Pan and Jingwei Hu for helpful discussions. CJ acknowledges the National Natural Science Foundation of China through grant 11873054, as well as the support by the Strategic Pioneer Program on Space Science, Chinese Academy of Sciences through grant XDA15052100. We acknowledge the science research grants from the China Manned Space Project with NO.CMS-CSST-2021-B11. This work is mainly based on the data from \xmm, an ESA science mission with instruments and contributions directly funded by ESA Member States and NASA. We acknowledge the use of public data from the \swift\ data archive. This research has made use of the NASA/IPAC Extragalactic Database (NED) which is operated by the Jet Propulsion Laboratory, California Institute of Technology, under contract with the National Aeronautics and Space Administration.\\

%

\vspace{5mm}
\facilities{{\it XMM-Newton}(EPIC), {\it Swift}(XRT)}


\software{
{\sc heasoft} (v6.28; \citealt{Blackburn.1995}), 
{\sc sas} (v18.0.0; \citealt{Gabriel.2004}),  
{\sc ds9} (v7.5; \citealt{Joye.2003}),
{\sc xspec} (v12.11.0m; \citealt{Arnaud.1996})
}
\\




\appendix
\section{Long-term Evolution of The Spectral-timing Properties of \sw16}
\label{app-sec-plots}
In this section we present the results of individual \xmm\ observations of \sw16, including the X-ray spectra and their best-fit absorbed power law models, the 0.3-1 keV and 2-10 keV light curves, flux distributions, PSDs and their best-fit power law models, the time-lag and coherence spectra using 2-10 keV as the reference band.

\begin{figure*}
\centering
\includegraphics[trim=0.0in 0.0in 0in -0.3in, clip=1, scale=0.65]{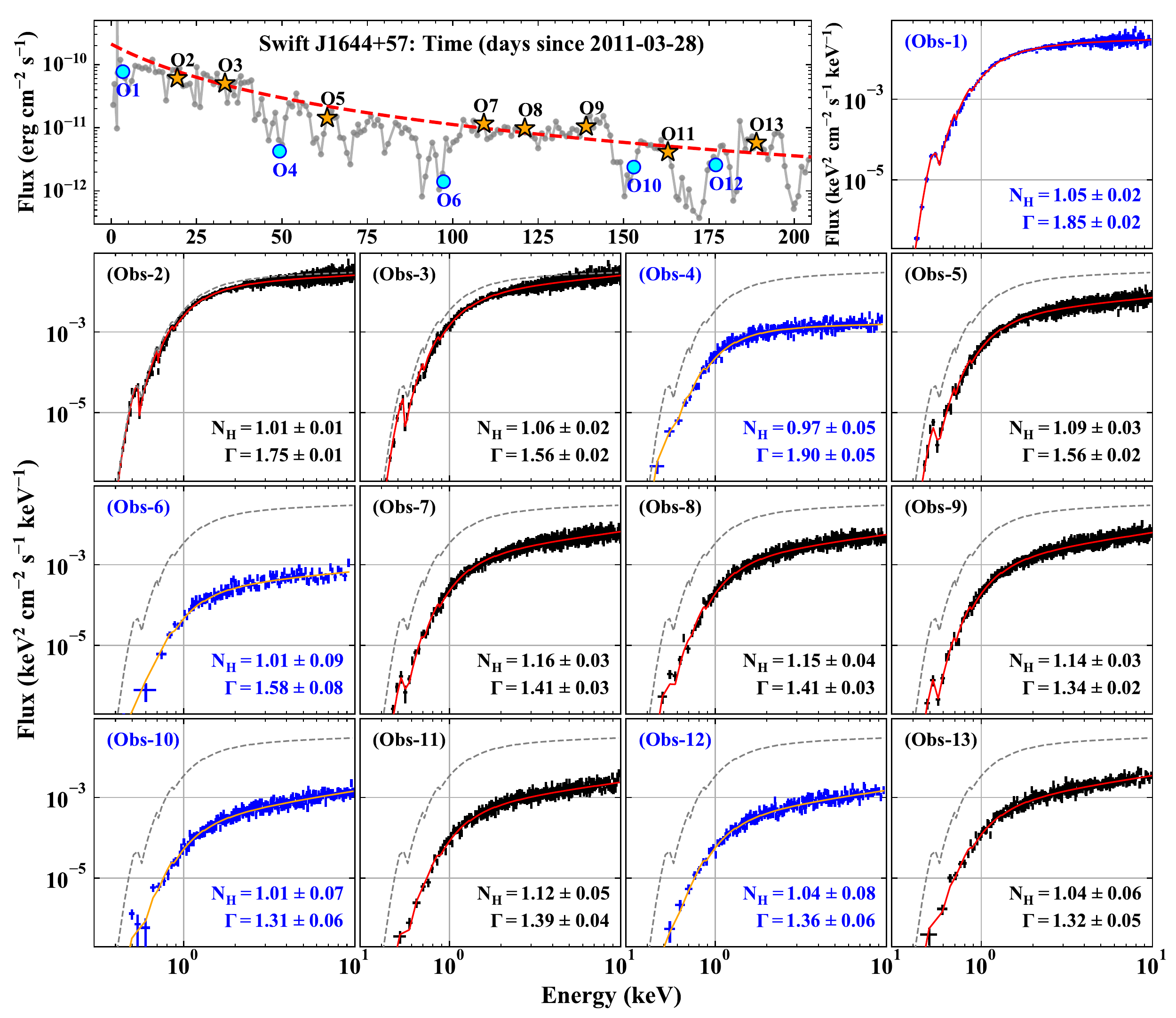}\\
\caption{The spectral evolution of \sw16\ as observed by \xmm/EPIC. The long-term light curve in the upper left panel is the same as Figure~\ref{fig-evo-spec}a. The 13 \xmm\ observations are divided into 2 groups according to the flux states, including normal state (black) and dipping state (blue). The following panels present the spectra. The red solid line shows the best-fit absorbed power law model. The gray dash line indicates the best-fit spectral model for Obs-1, plotted in other panels for comparison. The labelled $\Gamma$ is the 0.3-10 keV photon index. $N_{\rm H}$ is in units of $10^{22}$ cm$^{-2}$.}
\label{app-fig-spec}
\end{figure*}

\begin{figure*}
\centering
\includegraphics[trim=0.0in 0.0in 0in -0.3in, clip=1, scale=0.65]{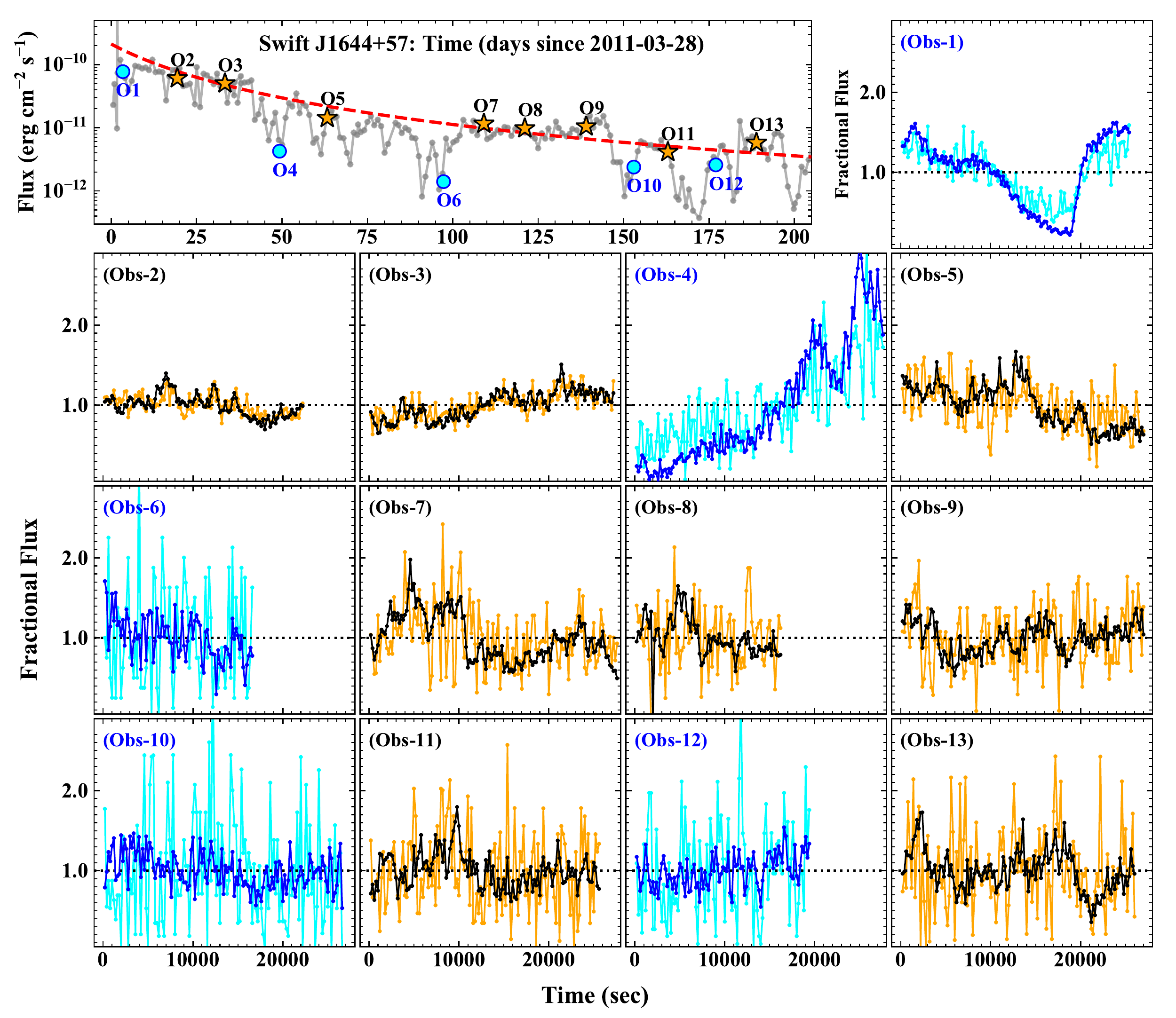}
\caption{The evolution of the X-ray light curve of \sw16\ as observed by \xmm/EPIC. The black and orange colors indicate the two energy bands of 2-10 keV and 0.3-1 keV in the normal state. The blue and cyan colors also indicate the two energy bands, but in the dipping state.}
\label{app-fig-lc}
\end{figure*}

\begin{figure*}
\centering
\includegraphics[trim=0.0in 0.0in 0in -0.3in, clip=1, scale=0.65]{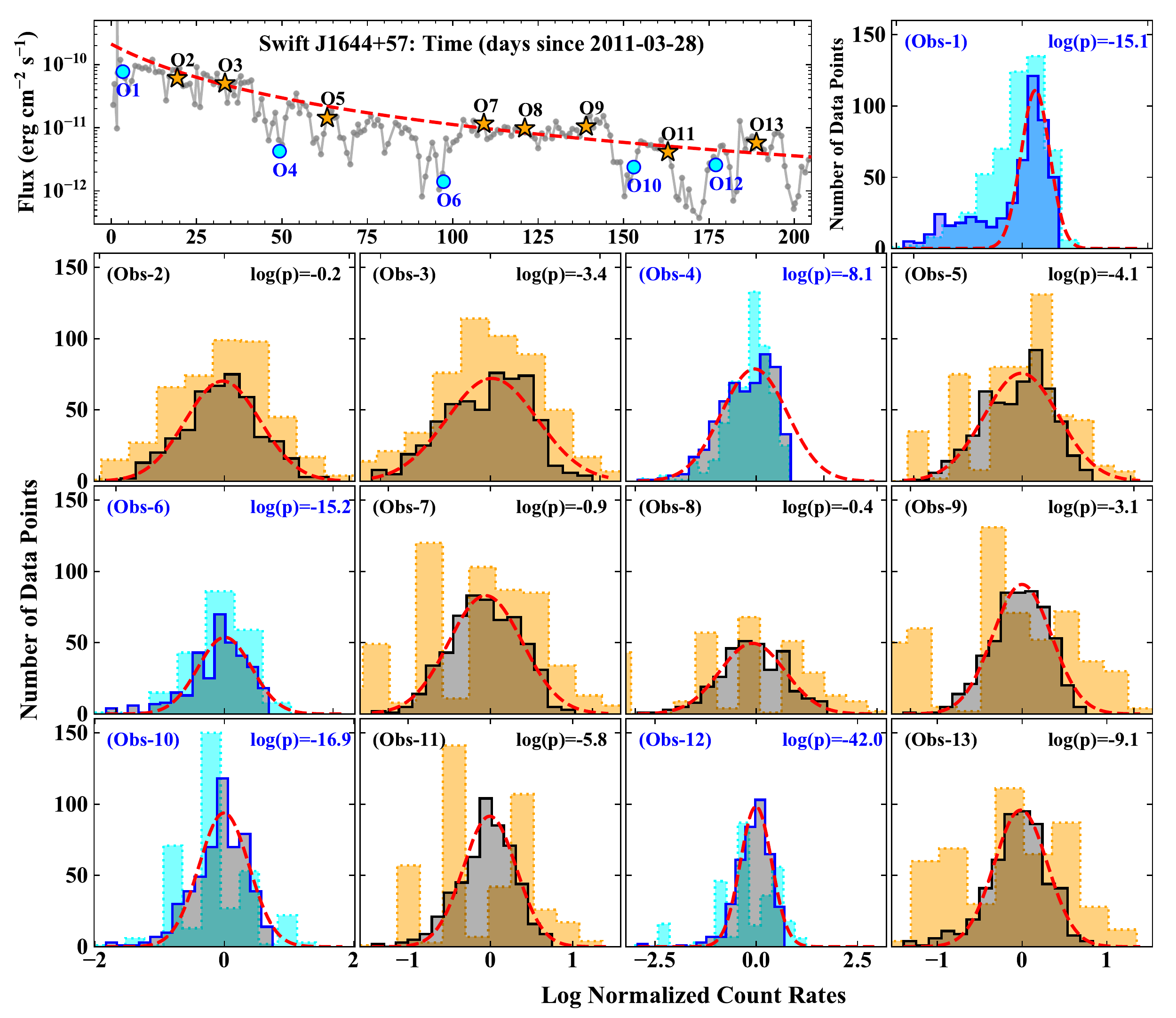}\\
\caption{The evolution of the short-term flux (count rate) distribution of \sw16, based on the light curves in Figure~\ref{app-fig-lc}. The back (blue) and orange (cyan) histograms represent the two energy bands of 2-10 keV and 0.3-1 keV. The red dash line indicates the best-fit normal distribution of the 2-10 keV count rate logarithms, and the labeled $p$-value indicates the deviation. These results show that the observed flux distribution in the dipping state deviates significantly from the lognormal distribution.}
\label{app-fig-ln}
\end{figure*}

\begin{figure*}
\centering
\includegraphics[trim=0.0in 0.0in 0in -0.3in, clip=1, scale=0.65]{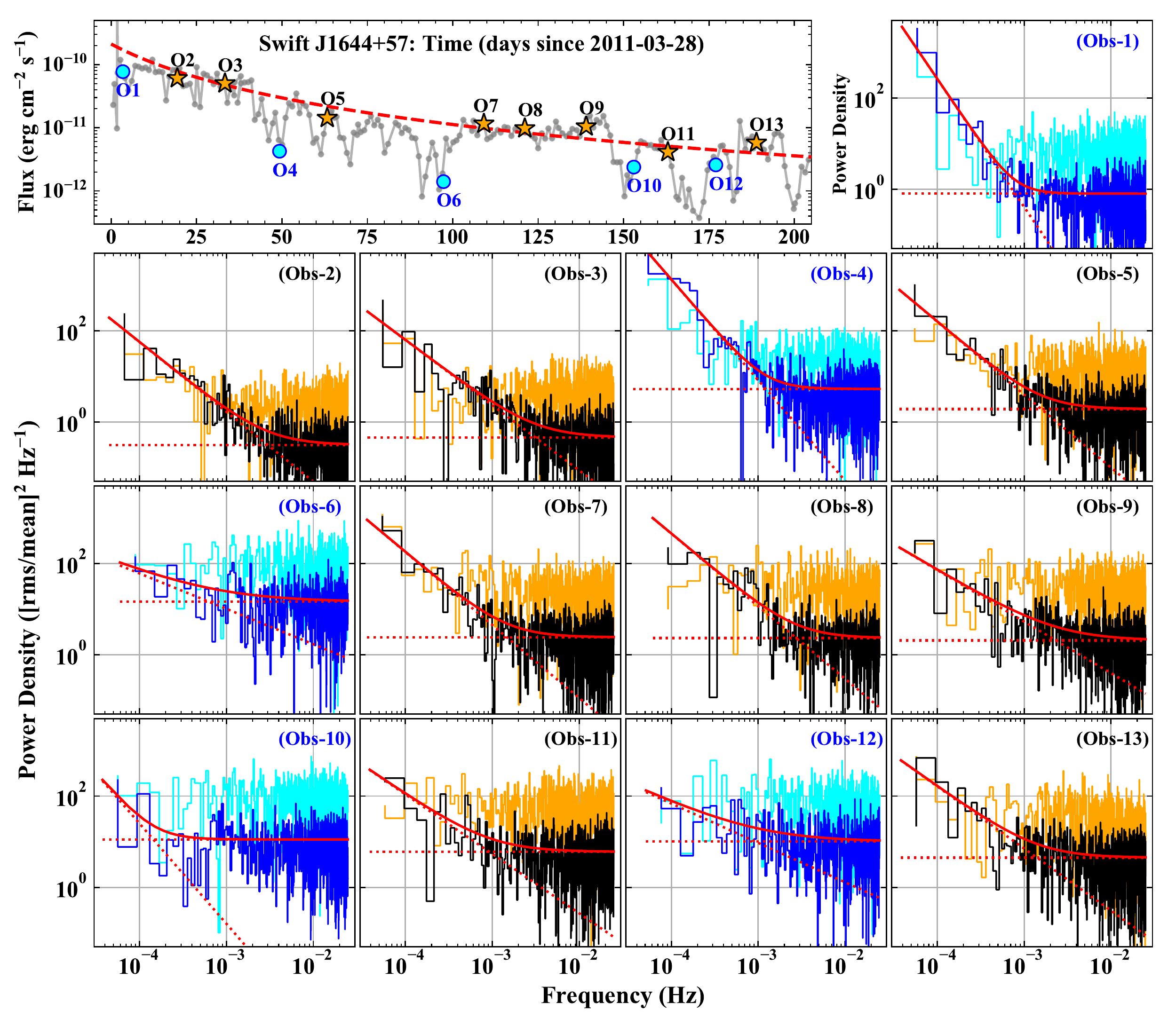}
\caption{The PSD evolution of \sw16, based on the light curves in Figure~\ref{app-fig-lc}. The red solid line indicates the best-fit model of the 2-10 keV PSD, which comprises a power law for the red noise continuum and a free constant for the Poisson noise, as indicated by the two red dotted lines.}
\label{app-fig-psd}
\end{figure*}

\begin{figure*}
\centering
\includegraphics[trim=0.0in 0.0in 0in -0.3in, clip=1, scale=0.65]{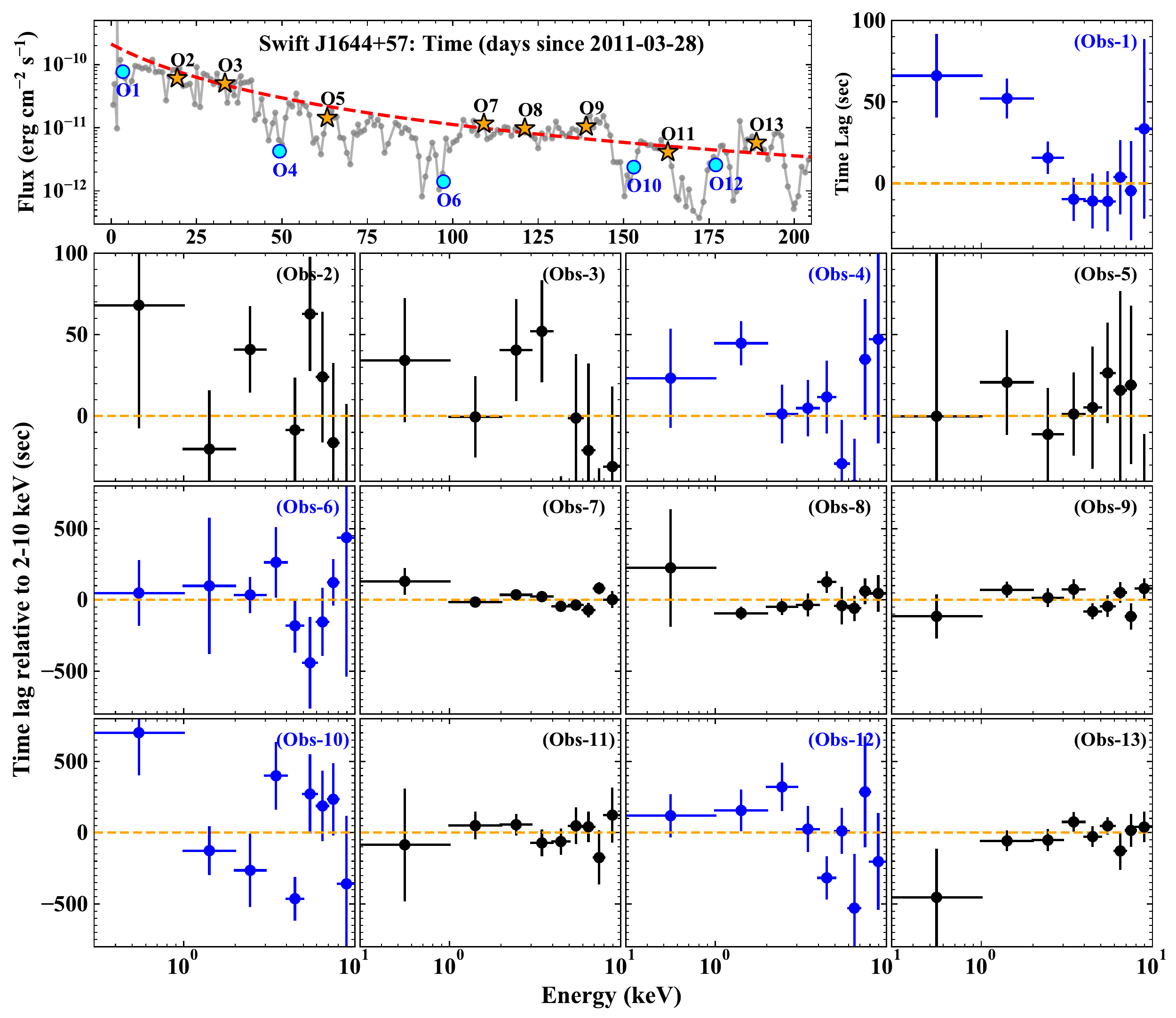}
\caption{The evolution of the lag spectrum of \sw16. A lag spectrum is calculated for every \xmm\ observation. The reference band is 2-10 keV, and the frequency range is between $(0.4-5)\times10^{-4}$ Hz (i.e. from 2 ks to the length of each light curve). A positive lag indicates that the variability in 2-10 keV lags behind. Significant positive lags are detected in Obs-1 and 4 in soft X-rays.}
\label{app-fig-lag}
\end{figure*}

\begin{figure*}
\centering
\includegraphics[trim=0.0in 0.0in 0in -0.3in, clip=1, scale=0.65]{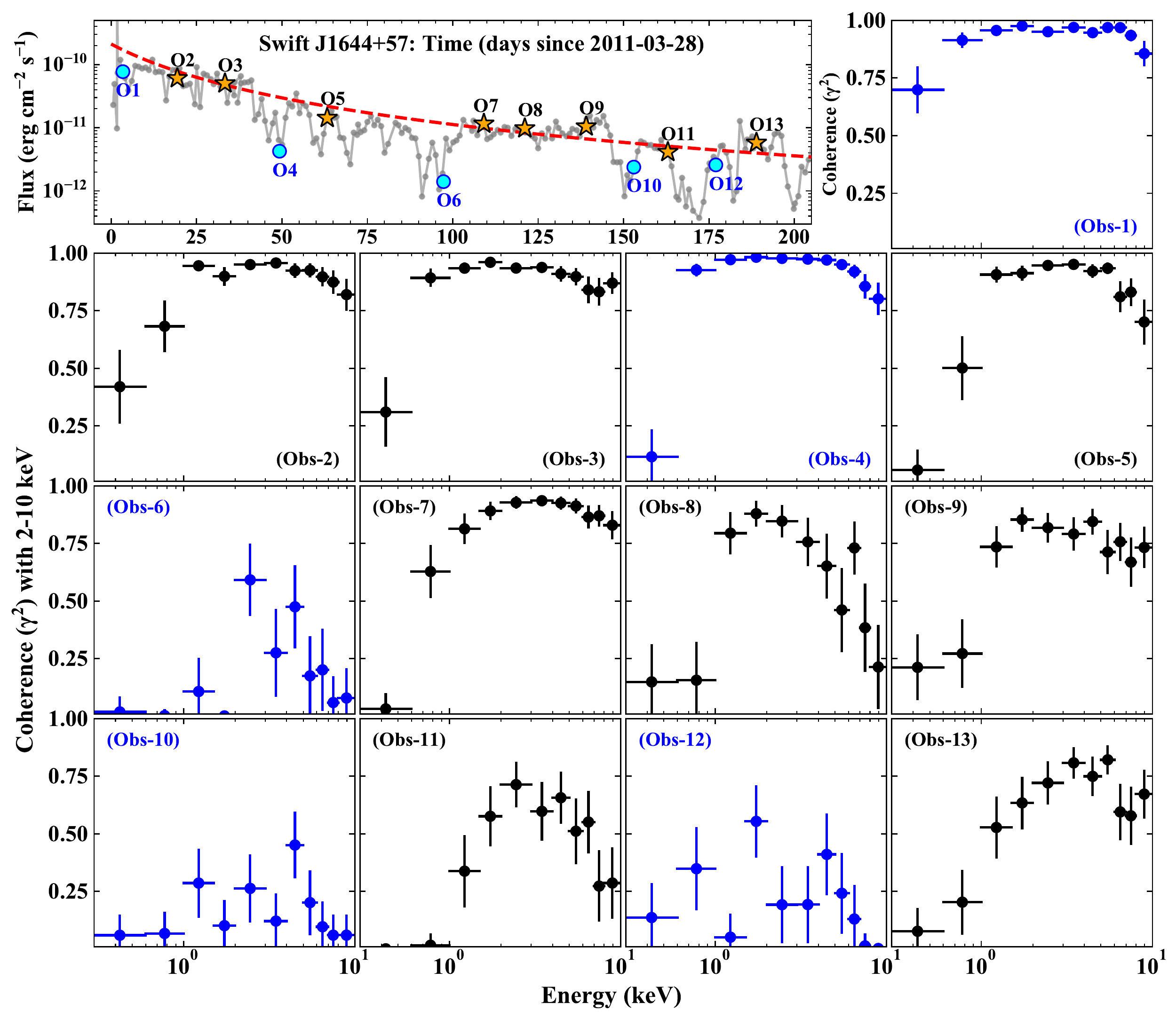}
\caption{The evolution of the coherence spectrum of \sw16. The reference band is 2-10 keV. The frequency range used is the same as for the lag spectra in Figure~\ref{app-fig-lag}.}
\label{app-fig-coh}
\end{figure*}





\end{document}